# THz near-field imaging of extreme subwavelength metal structures

Xinzhong Chen[1,±], Xiao Liu[2,±], Xiangdong Guo[3], Shu Chen[4], Hai Hu[3], Elizaveta Nikulina[4], Ziheng Yao[1,7], Hans Betchel[7], Michael C. Martin[7], G. L. Carr[8], Qing Dai[3*], Songlin Zhuang[2], Qing Hu[9], Yiming Zhu[2*], Rainer Hillenbrand[5,6], Mengkun Liu[1,8*], Guanjun You[2*]

[1]*Department of Physics and Astronomy, Stony Brook University, Stony Brook, New York 11794, USA*

[2]*Shanghai Key Lab of Modern Optical Systems, Terahertz Technology Innovation Research Institute, and Engineering Research Center of Optical Instrument and System, Ministry of Education, University of Shanghai for Science and Technology, Shanghai 200093, China*

[3]*Division of Nanophotonics, CAS Center for Excellence in Nanoscience, National Center for Nanoscience and Technology, Beijing 100190, China*

[4]*CIC nanoGUNE, 20018 Donostia-San Sebastián, Spain*

[5]*CIC nanoGUNE and UPV/EHU, 20018 Donostia-San Sebastián, Spain*

[6]*IKERBARSQUE, Basque Foundation of Science, 48013 Bilbao, Spain*

[7]*Advanced Light Source Division, Lawrence Berkeley National Laboratory, Berkeley, CA 94720, USA*

[8]*National Synchrotron Light Source II, Brookhaven National Laboratory, Upton, New York 11973, USA*

[9]*Department of Electrical Engineering and Computer Science and Research Laboratory of Electronics, Massachusetts Institute of Technology, Cambridge, Massachusetts 02139, USA*

[±]Those authors contribute equally.
[*]mengkun.liu@stonybrook.edu
[*]ymzhu@usst.edu.cn
[*]gjyou@usst.edu.cn
[*]daiq@nanoctr.cn

**Abstract:** Modern scattering-type scanning near-field optical microscopy (s-SNOM) has become an indispensable tool in material research. However, as the s-SNOM technique marches into the far-infrared (IR) and terahertz (THz) regimes, emerging experiments sometimes produce puzzling results. For example, "anomalies" in the near-field optical contrast have been widely reported. In this letter, we systematically investigate a series of extreme subwavelength metallic nanostructures via s-SNOM near-field imaging in the GHz to THz frequency range. We find that the near-field material contrast is greatly impacted by the lateral size of the nanostructure, while the spatial resolution is practically independent of it. The contrast is also strongly affected by the connectivity of the metallic structures to a larger metallic "ground plane". The observed effect can be largely explained by a quasi-electrostatic analysis. We also compare the THz s-SNOM results to those of the mid-IR regime, where the size-dependence becomes significant only for smaller structures. Our results reveal that the quantitative analysis of the near-field optical material contrasts in the long-wavelength regime requires a careful assessment of the size and configuration of metallic (optically conductive) structures.

## 1. Introduction

The best spatial resolution that can possibly be achieved with a conventional optical microscope is fundamentally constrained by the diffraction limit to roughly half of the incident wavelength, prohibiting nanoscale optical characterization [1]. In an effort to overcome this limit, the scattering-type scanning near-field optical microscope (s-SNOM) was developed. The tip-scattered light conveys local sample information and is detected by a conventional far-field detector. In this case, the spatial resolution is determined predominantly by the tip apex radius, tapping amplitude, and demodulation order, and as such is essentially independent of the incident wavelength [2]. Since its introduction, s-SNOM nano-imaging and spectroscopy have aided numerous scientific advances such as direct spatial mapping of surface polaritons, spatial phase coexistence, and electromagnetic field localization [3–11].

The spectral region spanning 0.1-10 terahertz (THz) is commonly recognized as the "THz gap" because of technical difficulties in the generation and detection of intense coherent THz light [12]. Performing s-SNOM at THz frequencies poses even greater challenges: conventional AFM probes are extreme subwavelength objects, resulting in low coupling and scattering efficiency. Nevertheless, intense research efforts by the THz s-SNOM community have led to significant technical and instrumental advances in recent years [13–26], enabling nano-imaging in the 50 GHz to 5 THz spectral region for accessing local conductivity and phase coexistence.

The deep-subwavelength scale spatial resolution achievable via s-SNOM can be largely attributed to the tight electromagnetic field confinement below the tip apex [27,28]. This fact sometimes leads to the misconception that only the small sample volume directly below the apex contributes to the detected scattering signal and that the optical properties, i.e. dielectric function or optical conductivity, of the material are the sole factors governing the near-field optical material contrast. However, recent THz s-SNOM experiments provide some contradicting and counterintuitive yet fundamentally interesting observations. For example, it has been reported that isolated metallic structures exhibit a lower near-field contrast compared to large and connected ones, even though the spatial resolution of THz s-SNOM is much higher than the structure size[11]. It has also been reported that AFM tips with larger apex radii increase the THz near-field amplitude and contrast, while leaving the spatial resolution nearly unaffected [28]. In this letter, we describe systematic THz and mid-IR near-field imaging of a series of metallic nanostructures. These findings suggest that in addition to intrinsic material properties, the size and configuration of the probed nanostructure in the deep sub-wavelength regime also plays important roles for near-field optical material contrasts.

## 2. Results and analysis
*2.1. Sub-THz nano-imaging*

We first use a home-built s-SNOM setup to perform sub-THz near-field imaging. The system utilizes high-harmonic (12th) generation from a microwave source [15,25]. A signal generator produces a microwave signal (9 GHz to 14.6 GHz), which is then frequency multiplied through Schottky diodes. This generates a coherent sub-THz wave (110 GHz to 175 GHz), which is directly focused onto the tip of an atomic force microscope (AFM) (Bruker Multimode). The AFM is operating at tapping mode with frequency Ω. The second signal generator supplies microwave radiation at a slightly different frequency of another Schottky diode, which functions as both the detector and the reference light generator. The frequency difference between the incident light and the reference light is set to be $\Delta\Omega = 24$ MHz. The second Schottky diode is connected to a Lock-in amplifier (Zurich HF2LI), which demodulates the signal at $n\Omega - m\Delta\Omega$ as an mean of phase-resolved imaging and far-field background suppression [29]. To improve the light coupling efficiency between the THz wave and the AFM tip [27,28], we use customized tips (Rocky Mountain Nanotechnology, LLC) of length about 200 μm and apex radius about 100 nm. The tips are longer and have a larger apex radius compared to typical AFM tips used for IR or visible s-SNOM, thus greatly enhancing the THz scattering and improving the signal-to-noise ratio [28]. All of our THz s-SNOM experiments use a typical tapping frequency of Ω ~40 kHz with a tapping amplitude around 100 nm.

We first characterize the spatial resolution of our THz s-SNOM setup via imaging a ~90-nm-thick by ~1.5 μm-wide gold bar antenna on a $SiO_2$ substrate. AFM topography and near-field amplitude images demodulated at the first (S1), second (S2), third (S3), and fourth (S4) harmonic orders are shown in Fig. 1(a). A clear material contrast between gold and $SiO_2$ is observed in all of the demodulated near-field amplitude images. Height and near-field optical signal profiles across the bar (designated by the white dashed lines) are displayed in Fig. 1(b). The near-field amplitude signals are normalized to the averaged substrate signal. The near-field optical resolution is similar to the AFM resolution and is mainly limited by the AFM tip apex radius. When a topographic boundary is present, it is difficult to precisely evaluate the optical resolution due to the unknown slope of the edge and edge-darkening artefact [30]. However, based on the edge profile of the optical signal, we conservatively estimate that the resolutions for AFM and S2 to be 120 nm and 200 nm based on 90%-10% signal width, similar to previously reported THz s-SNOM resolution [31]. Nonetheless, the achieved near-field resolution is in the order of $\lambda/10^4$, which is unprecedented at ~100 GHz range. Note that the relative contrast in the S1 image is rather weak due to an undesirable far-field background signal, which is confirmed by the approach curves in Fig. 1(c), wherein the S1 approach curve does not significantly subside even at a tip-sample distance of 1 μm, while the S2-S4 curves quickly decay. Similar to typical IR s-SNOMs, demodulation up to at least the second harmonic order is necessary in order to effectively suppress the background. Furthermore, statistical analysis of the signal variation on gold indicates that the S2 signal-to-noise ratio is at least 10:1.

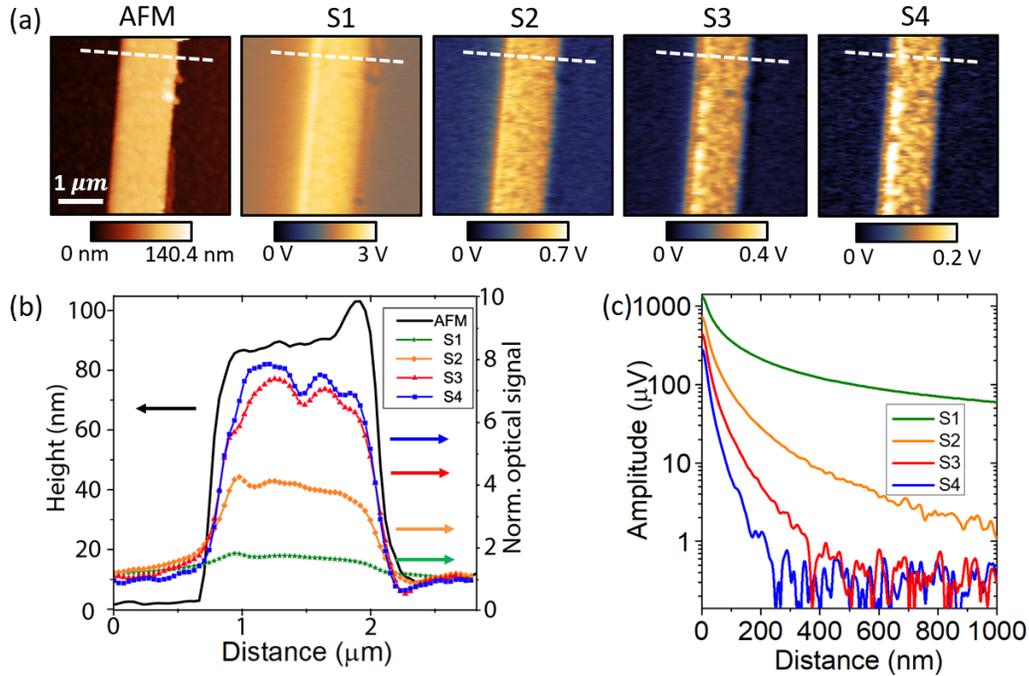

Fig. 1. Assessment of spatial resolution and signal-to-noise ratio obtained via THz s-SNOM. (a) Left to right: AFM topographic image and near-field images demodulated at different harmonics of the tip's tapping frequency. (b) Height and near-field optical signal profiles along the white dashed lines in (a). S1 through S4 are normalized to the average substrate signal in their corresponding images to yield a relative contrast. (c) Approach curves for different orders of demodulation on gold bar.

Next we demonstrate that the near-field material contrast of s-SNOM depends strongly on the size of the metallic structures, although the spatial resolution is nearly independent of the structure size. We first study gold disks of diameters from 1 to 5 μm, fabricated on a $SiO_2$ substrate (AFM image shown in Fig. 2(a)). From common knowledge one could assume an identical near-field material contrast for all disks, except for cases in which antenna effects or polaritonic responses occur (which is not the case here, as disks are of deep subwavelength scale). However, the THz s-SNOM amplitude images show a clear decrease of the near-field amplitude signal of the disk with decreasing disk diameter (Fig. 2(b)). Line profiles across the disk centers (Fig. 2(c)) show the effect more clearly. Note that even the smallest disk size (1 μm diameter) is much larger than the spatial resolution. We only display the S2 images without loss of generality. The S3 and S4 images show a similar trend.

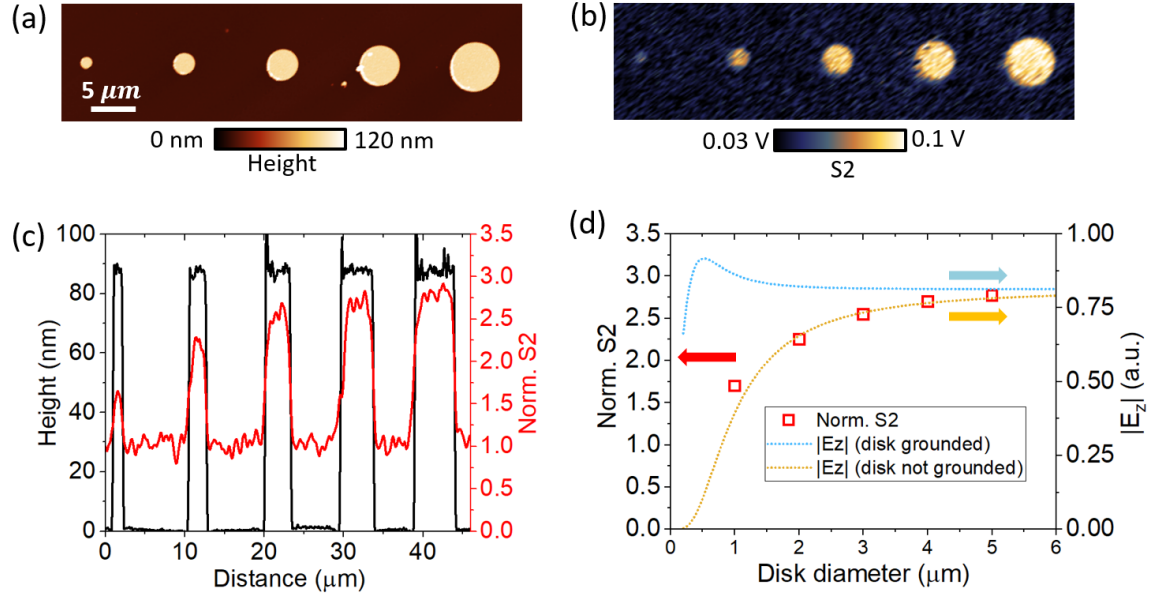

Fig. 2. (a) AFM image of a series of micron-sized gold disks. (b) The corresponding second harmonic near-field amplitude image at 171.84 GHz (1.74 mm), which shows an increasing near-field response with increasing disk diameter. (c) Height and normalized near-field S2 optical signal profiles across the disks in (a) and (b). The height of the disk is constant (~80 nm), while S2 increases with increasing disk diameter. (d) Red squares: average and normalized near-field signal (S2) of the disks with different diameters. Golden and blue curves represent the calculated electric field at the point charge for un-grounded and grounded disks, respectively.

Inspired by previous works on electrostatic force microscopy [32,33], we propose an electrostatic "toy-model" for our experimental data. The electrostatic limit is justified due to the much longer incident wavelength compared to the size of the tip and the nanostructures. For simplicity, we model the tip by an electrical monopole with charge $q$ located near the tip apex. Then the electrostatic problem of a point charge above a conducting disk can be solved analytically with a clean solution [34,35]. Suppose the point charge is located at $(0, 0, z_0)$ above the center of a grounded metal disk of radius $a$ in the $x - y$ plane, the potential on the z axis due to both the point charge and the induced surface charge on the disk is given by [35]:

$$V_1(z) = \frac{2q}{\pi(z^2 - z_0^2)} [z_0 \tan^{-1}\left(\frac{a}{z}\right) - z \tan^{-1}\left(\frac{a}{z_0}\right)]. \tag{1}$$

If the disk is not grounded, its potential picks up an extra term due to the presence of balancing charges with opposite sign (to make sure the disk is charge neutral) and becomes

$$V_2(z) = V_1(z) + \frac{2q}{a\pi} \tan^{-1}\left(\frac{a}{z}\right) \tan^{-1}\left(\frac{a}{z_0}\right). \tag{2}$$

In Fig. 2 (d), we calculate the trend of electric field at the position right below the monopole $E(z_0 \approx 250 \; nm) = \left.\frac{dV(z)}{dz}\right|_{z=z_0}$ as a function of disk diameter $d$ for both grounded (blue curve) and un-grounded (golden curve) scenarios. The balancing charge in the case of un-grounded disks is found to be of significant importance for the change of the electric field, which is consistent with the size dependent trend of the averaged S2 signal of the disks (red squares in Fig. 2(d)). On the other hand, when the disk is grounded, the electric field is predicted to be constant at $d > 1$ µm, as we demonstrate below.

*2.2. THz nano-imaging*

As suggested by equation 1 and the blue curve in Fig. 2(d), if the condition of local charge neutrality is broken, the near-field contrast will be boosted for small disks connected to a larger ground plane. To test this effect, nano-imaging is performed on two identical gold square patches with one isolated and the other connected to a large gold pad (Fig. 3). Here we use a commercial s-SNOM (Neaspec GmbH, Germany) with a tunable THz gas laser (SIFIR-50, Coherent Inc, USA) operating at 2.52 THz. The scattered light is collected interferometrically in a homodyne scheme by a cryogen-free THz bolometer (QMC Instruments Ltd., Cardiff, U.K.). Simultaneously obtained AFM topography and fourth harmonic near-field amplitude image are shown in Fig. 3(a) and (b). Corresponding line profiles across the gold patches and large gold pad are shown in Fig. 3(c). Despite the fact that two gold patches have the same thickness ~50 nm and lateral size ~1 µm, the near-field responses are drastically different. The connected gold patch exhibits a significantly stronger near-field signal (S4) than the isolated one, even slightly stronger than the large gold pad. This observation is consistent with the expectation and supports the validity of our electrostatic analysis.

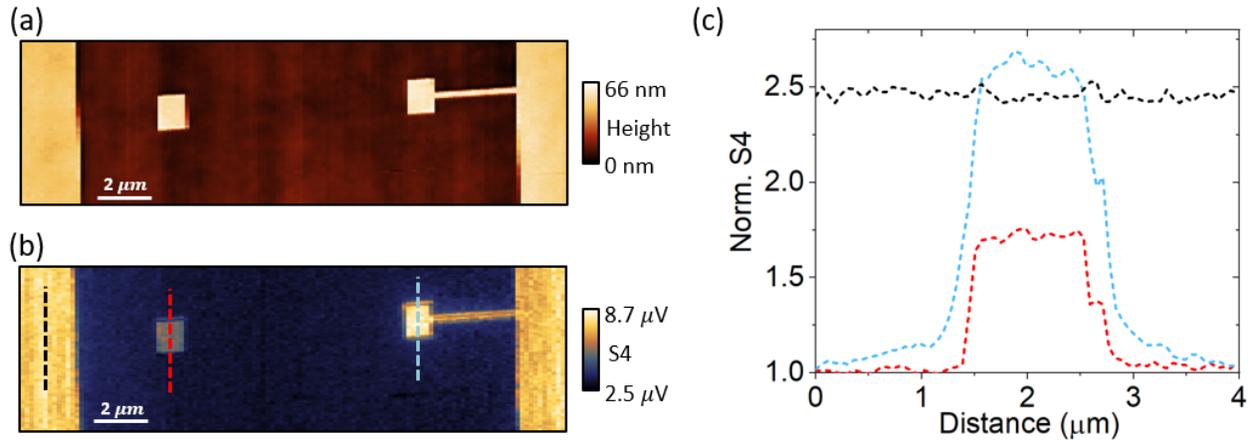

Fig. 3 (a) AFM topography image of an isolated gold square patch and a connected gold square patch. (b) The corresponding fourth harmonic near-field image (S4) at 2.52 THz (119 µm). (c) Normalized near-field S4 profiles along the indicated dashed lines in (b). The black curve is across the large gold pad. The blue curve is across the connected gold patch. The red curve is across the isolated gold patch.

Overall, our phenomenological model yields a qualitative interpretation of the observed phenomenon. However, we postulate that the underlying physics can be much more complex. Besides the above described effect, other effects may be responsible for the observed size-dependent near-field contrast which deserves further investigation. For example, beyond the electrostatic limit, surface inductance is found to be weakened in smaller metallic structures. The decreased inductance leads to a relatively more important contribution from the disk resistance, which leads to a higher eddy current damping in the near-field. This may greatly reduce the electric field intensity thus the near-field scattering signal (this analysis will be reported by S. T. Chui elsewhere).

*2.3. Full-wave numerical simulations*

While a full analytical electrodynamic model remains challenging, full-wave simulations of the tip-disk-substrate system are more probable and can be carried out using the finite element method (COMSOL). The scale mismatch among incident wavelength (1.74 mm), tip length (200 μm), and tip-sample distance (0-100 nm) makes a rigorous simulation of the near-field signal computationally expensive [28,30,36,37]. Instead, we use the electric field enhancement factor (electric field normalized to incident field) inside the tip-sample gap as an indirect gauge for the near-field interaction, a well-established approach within the community. It was recently demonstrated that the local field enhancement does not necessarily reflect the scattering intensity when the tip radii are different [28]. For example, enlarging a tip radius results in weaker local field enhancement even though it also makes the scattering intensity much higher. However, for a constant tip radius, the field enhancement and detected near-field signal are expected to be positively correlated [27]. Here we simulate the total electric field enhancement under the tip volume by modeling the tip as a cone of 200 μm shank length and 100 nm radius at its bottom, which resembles realistic tip geometry. The disk on the $SiO_2$ substrate is set to be 100 nm beneath the tip apex. An incident plane wave of 150 GHz frequency approaches the sample at a 60º angle with respect to the surface normal.

The simulation layout and typical electric field distribution are shown in Fig. 4(a). The field inside the gap between the tip and the disk is closely monitored as the disk size increases from 1 to 5 μm. To extend the generality of the simulation, we test three categories of material by adjusting the disk permittivity $\varepsilon = \varepsilon_1 + i\varepsilon_2$ for a good metal ($\varepsilon_1 \to -\infty, \varepsilon_2 \to \infty$), a dielectric material ($\varepsilon_1 = 10, \varepsilon_2 = 0.1$), and a bad metal ($\varepsilon_1 = -10, \varepsilon_2 = 10$). Field enhancement normalized to that of a "vacuum disk" ($\varepsilon_1 = 1, \varepsilon_2 = 0$) as a function of disk diameter is shown in Fig. 4(b). Of the three materials, only the good metal exhibits a significant, monotonically increasing field enhancement, as is consistent with our experimental observations. The field enhancements of the dielectric materials and bad metals are noticeably less affected by increasing disk diameter due to the lack of mobile charge carriers.

From the simulations, we conclude that the aforementioned size-dependent near-field contrast is most prominent in good metals, which have significant surface charge or current. The THz near-field signals of dielectrics or bad metals should not exhibit significant dependence for sample sizes at the micron scale (~$\lambda/100$).

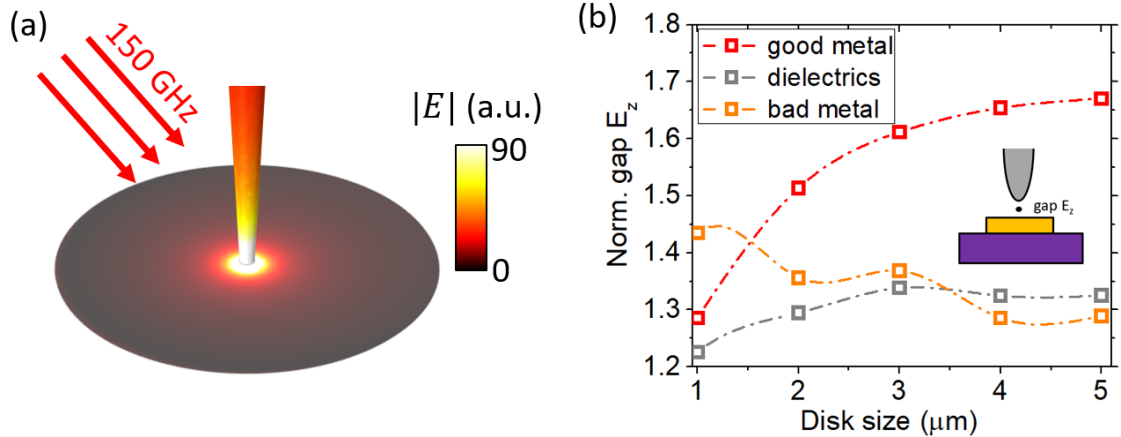

Fig. 4. (a) Simulation geometry and typical electric field distribution. (b) Normalized electric field in the tip-sample gap as a function of disk diameter for three categories of material: a good metal ($\varepsilon_1 \to -\infty$, $\varepsilon_2 \to \infty$), a dielectric material ($\varepsilon_1 = 10$, $\varepsilon_2 = 0.1$), and a bad metal ($\varepsilon_1 = -10$, $\varepsilon_2 = 10$).

*2.4. Mid-IR nano-imaging*

For comparison, we examine the size-dependent near-field material contrast in the mid-IR regime. Near-field imaging at ~6 μm is performed via a commercial s-SNOM (Neaspec GmbH, Germany) equipped with a quantum cascade laser (Daylight Solutions, USA). Fig. 5(a) shows S3 amplitude images of the same gold disks previously studied. Compared to the THz near-field images, no significant size-dependence is observed in the optical response of the mid-IR images for disks larger than 1 $\mu$m. Average near-field contrast as a function of disk size is plotted in Fig. 5(b) for two different tapping amplitudes. Increasing the tapping amplitude diminishes the overall contrast but has no significant effect on the trend of the optical response versus disk size. Unlike the sub-THz near-field signal, that of mid-IR quickly saturates when the disk size exceeds 1 μm. We further investigate by studying smaller disks of diameter 300 nm to 800 nm, whose responses quickly decay with decreasing disk size, as shown in Fig. 5(c).

To demonstrate that this phenomenon is not limited to a particular frequency, mid-IR broadband near-field spectroscopy is performed on the same disk samples. Indeed a similar trend is observed in the whole frequency range between 400 and 1100 cm$^{-1}$ (Supplemental materials). Furthermore, we find that demodulation order also has a minimal impact on trend of the near-field contrast (Supplemental materials).

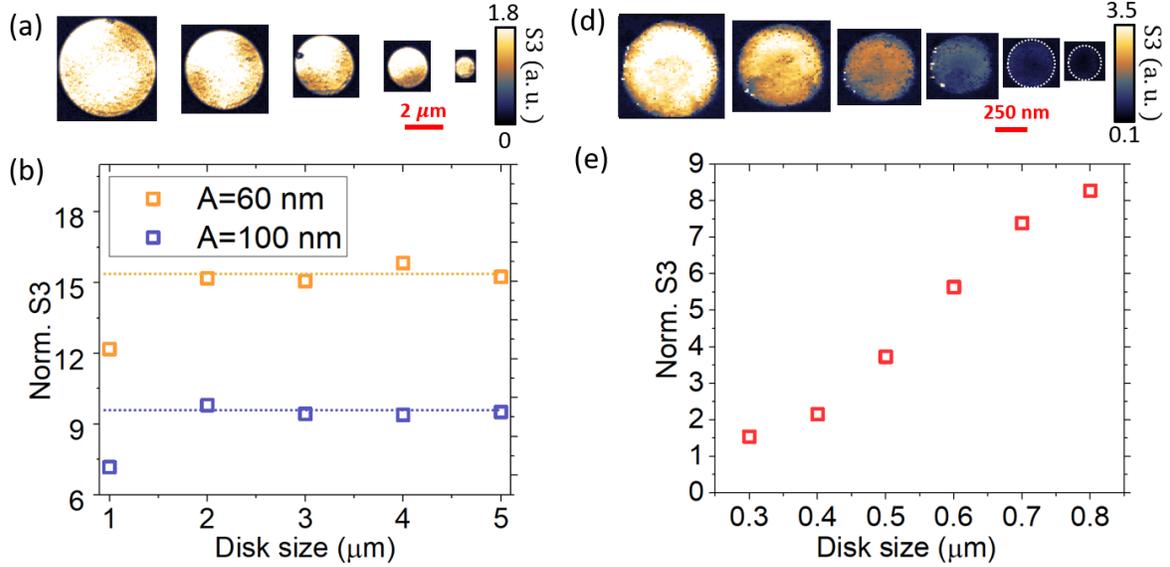

Fig. 5. (a) Near-field (S3) images on disks of diameter 1, 2, 3, 4, and 5 µm. (b) Average S3 contrast as a function of disk diameter. The dashed lines indicate the saturation signal level. A is the tip tapping amplitude. (c) Near-field (S3) images of disks of diameter 300, 400, 500, 600, 700, and 800 nm. White circles indicate the boundaries of the disks. (d) Average S3 contrast as a function of disk diameter. The dashed line indicates a linear trend.

## 3. Conclusion

Without considering polaritonic or antenna effects, current understanding of s-SNOM suggests that the near-field material contrasts are determined by the local dielectric properties of the sample and data acquisition parameters (tip radius, tapping amplitude, and demodulation order etc.). Therefore, theoretical attempts to calculate the near-field contrast typically model the sample as a homogeneous semi-infinite half-space [2,38–41] or layered planes [42,43], neglecting lateral inhomogeneity. These assumptions are largely valid for samples of bad metals or dielectric insulators [44,45]. However, our study shows that in metallic structures smaller than ~1/100 λ in the THz region and ~1/10 λ in the IR region, the sample size can significantly affect the near-field contrast and demands further theoretical investigations. Together with previous reports [46–48] demonstrating the IR near-field resonance effect in metallic structures of size ~1/10 λ to ~1 λ, this work suggests that geometric factors in metals are indeed important and need to be carefully addressed in the nano-imaging process with s-SNOM [49].

## Acknowledgment

This research is supported in part by the National Key Research and Development Program of China (Grant Nos. 2017YFF0106304, 2016YFF0200306), National Natural Science Foundation of China (61722111), 111 Project (D18014), and International Joint Lab Program supported by Science and Technology Commission Shanghai Municipality (17590750300). Stony Brook

University authors acknowledge support from National Science Foundation under Grant No. DMR-1904576. The CIC nanoGUNE authors acknowledge support from the Spanish Ministry of Economy, Industry, and Competitiveness (national project RTI2018-094830-B-100 and the project MDM-2016-0618 of the Marie de Maeztu Units of Excellence Program) and the H2020 FET OPEN project PETER (GA#767227). G.Y. thanks the Young Eastern Scholar at Shanghai Institutions of Higher Learning. This research used resources of the Advanced Light Source, which is a DOE Office of Science User Facility under contract no. DE-AC02-05CH11231. The authors would like to thank Dr. Oleg Butyaev (NT-MDT), Prof. S. T. Chui (University of Delaware), Dr. Alexander Mcleod (Columbia University), and Prof. D. N. Basov (Columbia University) for the helpful discussions.

**Notes**

The authors declare the following competing financial interest(s): R.H. is a co-founder of Neaspec GmbH, a company producing scattering-type scanning near-field optical microscope systems such as the one used in this study. The other authors declare no competing interests.

**Supplemental Materials**

**THz nano-imaging on linear antennas with different lengths**

Similar to the disks, a system of rectangular gold bars exhibits similar size-dependent S2 signals. The gold bars are 55 nm thick and 1 μm wide, and 1.5 μm, 2.5 μm, 5.5 μm, and 10.5 μm in length. AFM and near-field images are shown in Figs. S1(a) and (b), while Fig. S1(c) shows the height and near-field optical signal profiles along the long axes of the gold bars. Size-dependent S2 behavior is clearly observable.

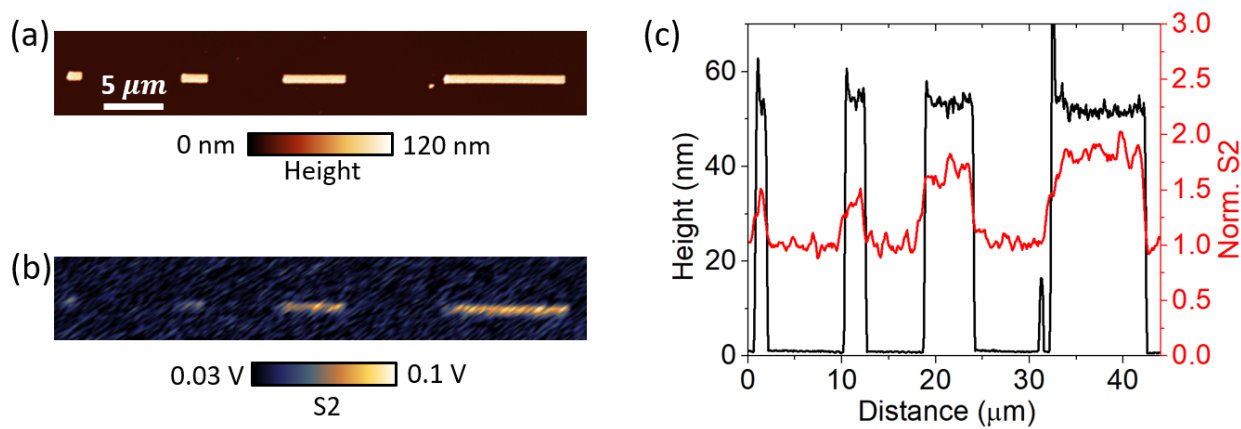

Fig. S1. (a) AFM image of a series of micron-sized gold bars. (b) The corresponding second harmonic near-field image at 171.84 GHz (1.74 mm), which shows an increasing near-field response with increasing bar length. (c) Height and near-field S2 optical signal profiles across the long axes of the bars in (e) and (f).

**Mid-IR broadband spectroscopy on gold disks**

In addition to nano-imaging, broadband spectroscopy is performed on disks with diameters of 1 to 5 μm in the mid-IR frequency range. Near-field spectra are taken when the tip is landed at the center of each disk and shown in Fig. S2. The spectral features at ~475 cm$^{-1}$ and ~1100 cm$^{-1}$ (indicated by red arrows) are due to the IR active phonon modes of the $SiO_2$ substrate. With increasing disk size, the spectrum is generally elevated, gradually screening the phonon response of the substrate. This broadband Synchrotron Infrared Nanospectroscopy (SINS) experiment is performed at beamline 2.4 of Advanced Light Source at Lawrence Berkeley National Laboratory.

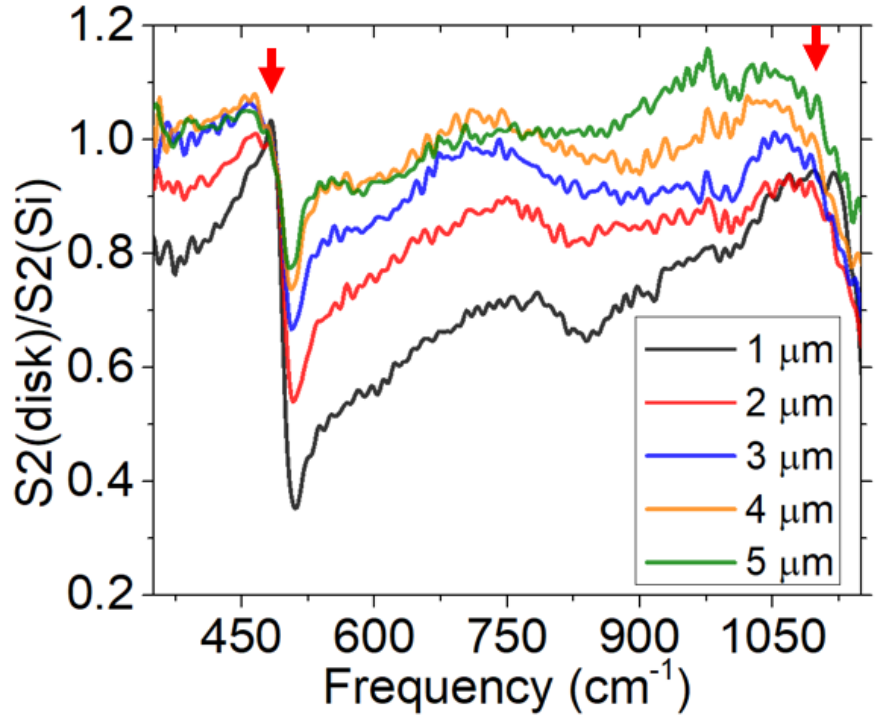

Fig. S2. Mid-IR broadband spectroscopy on 1-5 μm disks.

**Effect of demodulation order on near-field contrast**

Higher demodulation order generally leads to larger relative contrast due to the better confinement of the electric field "hot spot" under the tip. Compare Fig. S3 to Fig. 6(b), we found that the demodulation order has a minimal impact on the near-field contrast trend in our study here.

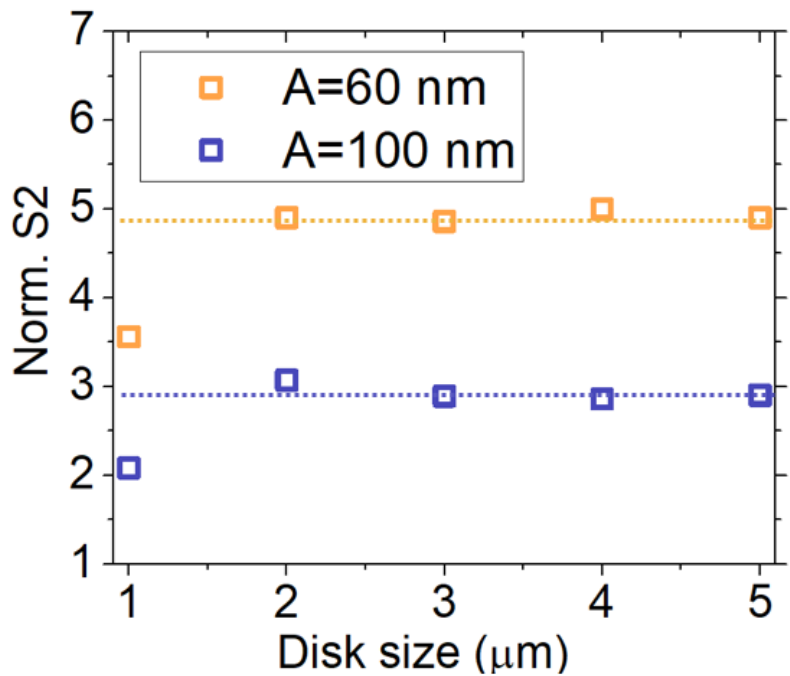

Fig. S3. Average S2 contrast as a function of disk diameter. The dashed lines indicate the saturation signal level. A is the tip tapping amplitude.